\documentclass[aps,twocolumn,showpacs,11pts]{revtex4}

\usepackage{amsmath}
\usepackage{mathrsfs}
\usepackage{graphicx}

\newcommand{\ket}[1]{|\hspace{0.5pt}#1\hspace{0.5pt}\rangle}
\renewcommand{\v}[1]{\ensuremath{\mathbf{#1}}}
\begin{document}

\title{Validity of the lowest-Landau-level approximation for rotating Bose
gases}


\author{Alexis G. Morris}
\author{David L. Feder}
\affiliation{Department of Physics and Astronomy, and Institute for Quantum
Information Science, University of Calgary, Calgary, Alberta, Canada T2N 1N4}

\begin{abstract}
The energy spectrum for an ultracold rotating Bose gas in a
harmonic trap is calculated exactly for small systems, allowing
the atoms to occupy  several Landau levels. Two vortex-like states
and two strongly correlated states (the Pfaffian and Laughlin) are
considered in detail. In particular, their critical rotation
frequencies and energy gaps are determined as a function of
particle number, interaction strength, and the number of Landau
levels occupied (up to three). For the vortex-like states, the
lowest-Landau-level (LLL) approximation is justified only if the
interaction strength  decreases with the number of particles;
nevertheless, the constant of proportionality increases rapidly
with the angular momentum per particle. For the strongly
correlated states, however, the interaction strength can
\emph{increase} with particle number without violating the LLL
condition. The results suggest that in large systems, the Pfaffian
and Laughlin states might be stabilized at rotation frequencies
below the centrifugal limit for sufficiently large interaction
strengths, with energy gaps a significant fraction of the trap
energy.
\end{abstract}

\pacs{03.75.Lm, 05.30.Jp, 73.43.-f}

\maketitle


\section{Introduction}\label{intro}

Over the past few years, there has been much interest in the possibility of
coaxing ultracold atomic gases into the neutral analog of a quantum Hall
state. Various approaches have been proposed, which include rapidly rotating
the confinement potential~\cite{Wilkin:1998,Wilkin:2000,Cooper:2001,
Paredes:2002,Reijnders:2002,Ho:2002,Regnault:2003,Reijnders:2003,Regnault:2004,
Cooper:2004,Popp:2004,Cooper:2005,Baranov:2005,Cazalilla:2005}, and generating
effective magnetic fields for particles confined in optical lattice
potentials~\cite{Jaksch:2003,Mueller:2004,Sorensen:2005,Osterloh:2005} and
harmonic oscillator traps~\cite{Ruseckas:2005,Ohberg:2005}.

One of the many intriguing prospects for these systems is the possibility of
producing the bosonic analog of the
Pfaffian~\cite{Wilkin:2000,Regnault:2003,Cooper:2004,Cazalilla:2005}, also
known as the Moore-Read state~\cite{Moore:1991}, which is believed to give
rise to the $\nu=5/2$ fractional quantum Hall plateau in two-dimensional
electron gases~\cite{Pan:1999}. The quasiparticle excitations of this state
are believed to be non-Abelian anyons obeying fractional
statistics~\cite{Sarma:2005,Bonderson:2005,Bonderson:2006}; in principle,
non-Abelian anyons could be physically wound around one another to operate an
intrinsically fault-tolerant quantum computer, protected from errors by the
topological properties of the underlying state~\cite{Kitaev:2003,
Bonesteel:2005,Sarma:2005}.

One of the fundamental sources of error in these topological approaches to
computing is the thermal excitation of unwanted quasiparticles that contribute
to the winding: the error rate is intrinsically related to the size of the
energy gap separating the ground state from the next excited one. One might
expect that strengthening the particle interactions would
monotonically increase the splitting between sub-bands in the Landau
levels; this would also allow the system to access a larger number of quantum
Hall states. Indeed, for ultracold atomic systems the size (and sign) of the
scattering length (characterizing the interaction strength) can be varied
arbitrarily through the use of Feshbach resonances~\cite{Cornish:2000}. It is
not known, however, how large interaction strengths affect the quantum Hall
states and the excitation gaps, except right on resonance~\cite{Cooper:2004}.

The difficulty is that calculations used to locate the various
quantum Hall ground states usually rely on the Lowest Landau Level
(LLL) approximation, which is valid only in the weakly interacting
regime. This is turn requires that the particle densities are low,
the interaction strength is small, or the rotation frequencies are
exceedingly close to the harmonic oscillator confinement
frequency. But robust quantum Hall states with large excitation
gaps require strong interactions.

In the present work, we exactly calculate the ground states for
rotating bosons for small systems without employing the LLL
approximation. The central goals are (1) to determine an upper
bound to the interaction strength consistent with the LLL
approximation; (2) to obtain the physical parameters (number of
bosons, scattering length, rotation frequency) to unambiguously
select a given quantum Hall state; and (3) to show that negative
(attractive) interactions always lead to instability for large
densities, contrary to prior
claims~\cite{Fischer:2004,Lakhoua:2005}. The main result of this
paper is that the widely accepted criterion for the LLL
approximation, which requires that the inter-particle interaction
energy be smaller than the Landau level spacing, is generally too
restrictive: we expect that the most interesting quantum Hall
states can be obtained for relatively large particle numbers and
rotation frequencies within reach of current experiments.

In Sec.~\ref{method}, we describe the Bose gas and our approach to
calculating the boson energy spectrum. Limits on the validity of
the LLL approximation are investigated in Sec.~\ref{positive-g} by
considering the effects of higher Landau levels on the boson
energy spectrum. The case of attractive Bose gases is discussed in
Sec.~\ref{negative-g}, and concluding remarks are provided in
Sec.~\ref{conclusion}.

\section{Numerical Method}\label{method}
The system under study consists of ultracold
interacting bosons subject to a cylindrically symmetric harmonic trapping
potential, which is rotated around the $z$-axis at a frequency
$\tilde{\Omega}$.
We assume a tight harmonic confinement along the axis of rotation such that
the axial ground state energy far exceeds any other transverse energy scale,
yielding a quasi-2D system.
The Hamiltonian in the co-rotating frame is then $H=H_0 + H_{\rm int}$, where
\begin{equation*}
H_0=\sum_i^N \left(\frac{P^2_i}{2M} + \frac{1}{2}M\omega^2r_i^2 -
\tilde{\Omega} L_{i}\right) \label{H0}
\end{equation*}
and
\begin{equation}
H_{\rm int}=\tilde{g}\sum_{i<j}^N \delta(\v{r}_i-\v{r}_j).
\label{Hint}
\end{equation}
Here, $M$ is the particle mass, $N$ the number of bosons, $\omega$ the radial
trap frequency and $\tilde{g}=\sqrt{8\pi}\hbar\omega \ell^2 a/\ell_z$ is the
2D-interaction strength where variables $\ell=\sqrt{\hbar/M\omega}$ and  $\ell_z=\sqrt{\hbar/M\omega_z}$ are the characteristic oscillator lengths
along the radial and axial directions respectively, and $a$ is the
three-dimensional scattering length~\cite{Petrov:2001}.

We proceed to calculate the energy spectrum of the rotating Bose gas
by exact diagonalization of the Hamiltonians $H_{\rm int}$ and $H_0$ in
blocks of definite total angular momentum.
To do so, we choose a Fock basis of the form
\begin{equation*}
\ket{{\cal N}_{1},{\cal N}_{2},\ldots}=\prod_\v{k}\frac{(\hat
b_\v{k}^{\dagger})^{{\cal N}_\v{k}}\ket{0}}{\sqrt{{\cal N}_\v{k}!}},
\label{Fock}
\end{equation*}
where $\hat b_\v{k}^{\dagger}$ creates a boson in state $\v{k}$
(in our case the index $\v{k}$ represents a distinct pair of
numbers $(n,m)$, the principle quantum number and the projection
of the angular momentum) and $\cal N_\v{k}$ is the occupation
number of level $\v{k}$. The Hamiltonians are then written in
terms of the Bose field operators where $\hat H_{0} = \int
\hat\psi^\dagger(\v{r}) H_0 \hat\psi(\v{r})d\v{r} $  and $ \hat
H_{\rm int} =  \frac{1}{2}\int \hat\psi^\dagger(\v{r})
\hat\psi^\dagger(\v{r'}) H_{\rm int} \hat\psi(\v{r'})
\hat\psi(\v{r}) d\v{r} d\v{r'} $.  Expanding the field operators
in terms of 2D harmonic oscillator basis states,
$\hat\psi^\dagger(\v{r}) = \sum_{\v{k}} \hat b^\dagger_{\v{k}}
\Phi_\v{k}(\v{r}) $, one obtains
\begin{equation*}
\hat H_0 =\sum_{\v{k}}\hat b^\dagger_{\v{k}}\hat b_{\v{k}} \epsilon_{\v{k}}
\end{equation*}
and
\begin{equation}
\hat H_{\rm int} = \frac{\tilde{g}}{2}\sum_{\v{ijkl}} \hat b_{\v{i}}^\dagger
\hat b_{\v{j}}^\dagger\hat b_{\v{k}}\hat b_{\v{l}}\int \Phi^*_{\v{i}}(\v{r})
\Phi^*_{\v{j}}(\v{r}) \Phi_{\v{k}}(\v{r}) \Phi_{\v{l}}(\v{r})
d\v{r}, \label{Hint-2nd}
\end{equation}
where the $\epsilon_{\v{k}} = \hbar\omega(2n+|m|+1 -m\Omega)$ are the
single particle 2D harmonic oscillator eigenvalues, and
$\Omega=\tilde{\Omega}/\omega$ is the dimensionless rotation frequency.

In our Fock basis, $H_0$ is diagonal with eigenvalues of $E_n/\hbar\omega
= 2\sum_i^N n_i+ \sum_i^N|m_i| + N -\Omega L, \label{E0}$ where $L=\sum_i^Nm_i$
is the total angular momentum of the system in units of $\hbar$. Thus, in the
purely non-interacting case, the energy spectrum consists solely of degenerate
levels separated by an energy gap $2\hbar\omega$.  We will adopt the
usual terminology where these different levels are called Landau Levels,
in reference to the similarities between the rotating Bose gas and
the 2D electron gas subjected to a strong perpendicular magnetic field.
The LLL approximation corresponds to enforcing that all particles have $m_i\ge0$ and $n_i = 0$, which greatly reduces the Hilbert
space dimension. The interaction Hamiltonian, on the other hand, is not
diagonal in our basis and so once it is included the degeneracy
of the Landau levels is lifted and each Landau level is split into a multitude of distinct states. The interaction strength is
parameterized by the dimensionless coupling constant
$g = \tilde{g}/\hbar\omega\ell^2 = \sqrt{8\pi} a/\ell_z$.

\begin{figure}[t]
    \begin{center}
    \includegraphics[width=0.4\textwidth]{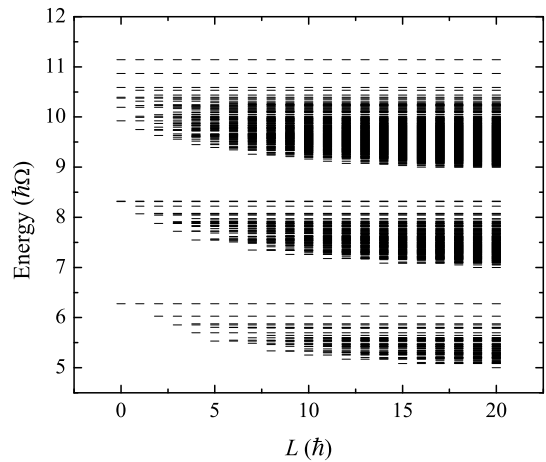} \caption{Energy
    spectrum as a function of total angular momentum for $N=5$ bosons in the Laughlin ground state, with three
    Landau Levels.  The rotation rate was set to $\Omega = 1$
    and the interaction strength to $g=1$.  States with $L<0$ are not considered
    in this work since these are of higher energy than those with $L>0$ (due to the energy eigenvalue Doppler shift term $-\Omega L$).} \label{spectrum_pos}
    \end{center}
\end{figure}

For small number of particles it is feasible to calculate the
entire spectrum, but for larger $N$ and greater number of Landau
Levels the Hilbert space becomes too large. In these cases, only
the lowest-lying eigenvalues were calculated through the use of a
Lanczos diagonalization algorithm. Even so, computational
constraints have limited our calculations to the first two or
three Landau Levels, which are obtained by including states where
$\sum_i^N \left[n_i + (|m_i| - m_i)/2\right] \le 1$ or 2,
respectively. The 5 particle spectrum with $g=1$ and $\Omega=1$ is
shown in Fig.~\ref{spectrum_pos}, including the three lowest
Landau levels.

\section{Repulsive Interactions}\label{positive-g}

We begin by considering a repulsive Bose gas.  As the interaction
energy is increased, mixing between the different Landau Levels
becomes more important and at some critical interaction strength
$g_{\rm max}$, the LLL approximation can no longer accurately
describe the rotating gas. The standard criterion for the validity
of the LLL approximation is that the interaction energy should be
smaller than the spacing between Landau Levels, or $\tilde{g}\rho
< 2\hbar\omega$ where $\rho$ is the particle density
\cite{Wilkin:1998, Mottelson:1999, Jackson:2000, Liu:2001,
Stock:2005} In unitless form, the crossover between the weakly and
strongly interacting regime occurs when
\begin{equation}
g_{\rm max}\sim N^{-1}.
\label{g_scaling_trad}
\end{equation}
This scaling does not provide an absolute estimate on $g_{\rm max}$, however;
the coefficient of proportionality might be large enough that interesting
quantum Hall states could be achieved for experimentally accessible rotation
frequencies and interaction strengths. Most important though, the $1/N$
scaling is found to be invalid for the most interesting ground states, as discussed below.

\begin{figure}[t]
    \includegraphics[width=0.48\textwidth]{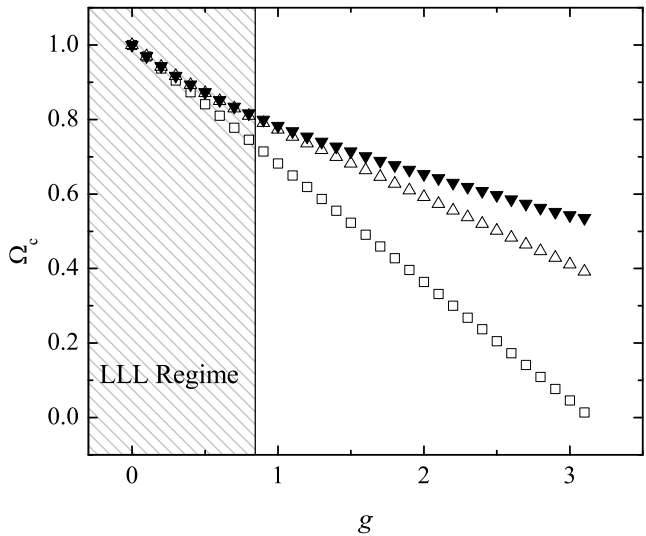}
    \caption{The critical rotation frequency $\Omega_c$ characterizing the transition
    from the $L=0$ to $L=N$ state for 8 particles, as a function of interaction
    strength $g$. The regime of validity of the LLL approximation is found by determining at what
    value of $g$ a difference of more than 10\% appears between $\Omega_c$ found using the LLL
    approximation and $\Omega_c$ found using more Landau Levels.  Squares,
    open triangles, and closed triangles depict the results for one (LLL
    approximation), two, and three Landau levels, respectively.  This same labeling
    scheme will be used in all subsequent figures.} \label{gmax}
\end{figure}

For a non-rotating trap, the stationary Bose gas with no angular momentum has
the lowest energy. However, as the rotation rate is increased, states with
higher $L$ experience a greater Doppler shift in energy than those with lower
$L$ and, as a result, at specific rotation frequencies $\Omega_c$ ground
state transitions are observed.  Almost all stable ground states can
be described by a relationship of the form $L=a(N-b)$, where $a$ and
$b$ are integers~\cite{Wilkin:2000}.  We will be concentrating on the
$L=N$ state (sometime called the single vortex state), the $L=2(N-1)$ state,
the Pfaffian (where $L=(N-1)^2/2$ for odd $N$ and $L=N(N-1)/2$ for even $N$),
and the Laughlin state for which $L=N(N-1)$.

\begin{figure}[t]
    \includegraphics[width=0.48\textwidth]{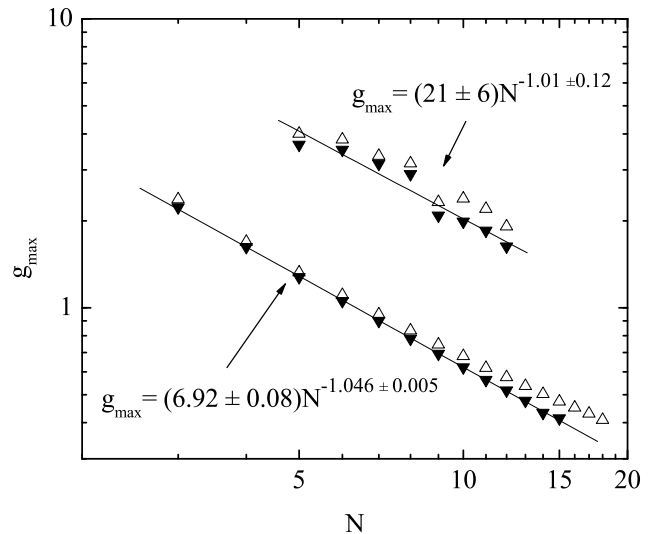} \caption{Scaling
    of $g_{\rm max}$ with number of particles for the $L=N$ (lower set) and
    $L=2(N-1)$ (upper set) states.  The $g_{\rm max}$ values were obtained
    using the method described in Fig.~\ref{gmax}. Open and filled triangles depict the
    results for two and three Landau levels, respectively. The straight lines
    are the best fits through the 3LL data.}
    \label{fits_weak} 
\end{figure}

To characterize the transition from the weakly interacting to strongly
interacting regime, we consider changes in the critical trap rotation
frequency $\Omega_c$ of these four ground states as higher Landau levels are
included in the eigenvalue calculations, as shown in Fig.~\ref{gmax}. When
the difference between $\Omega_c$ obtained with the LLL approximation and that
obtained with higher Landau levels in the spectrum become apparent, we say
that the LLL approximation is no longer applicable.  Specifically, we define
the crossover from weakly to strongly interacting regimes to occur when the
relative error between the results of the LLL approximation and those with two
or three Landau levels exceeds a threshold around 10\%:
\begin{equation*}
\frac{|\Omega_c(g) - \Omega_c(g)_{LLL}|}{\Omega_c(g)_{LLL}} = 0.1.
\label{criteria}
\end{equation*}
The interaction strength for which this occurs is defined as
$g_{\rm max}$, and depends on the number of particles $N$ and the
particular transition considered.  The criterion of 10\% was
chosen to reflect typical experimental uncertainty; we explicitly
considered error thresholds of 5\% and 15\%, and the results were
not significantly changed. We also repeated the entire calculation
using the gap $\Delta E$ between the ground and first excited
state as the marker; however, this method proved to be unreliable.
While ${g_{\text{max}}}$ values found using $\Delta E$ tend to be
smaller than those obtained with the $\Omega_c$ criterion, there
is too much scatter to obtain a useful relation between $g_{\rm
max}$ and $N$.  This is a consequence of the small number of
particles considered: as $N$ is increased, the energy gap
undergoes considerable fluctuations as new states are interjected
into the spectrum. Furthermore, it is more valuable from an
experimental perspective to consider variations in $\Omega_c$
since the trap rotation frequency can be directly controlled,
whereas $\Delta E$ is indirectly controlled through the choice of
$g$ and $\Omega$.

\begin{figure}
    \includegraphics[width=0.48\textwidth]{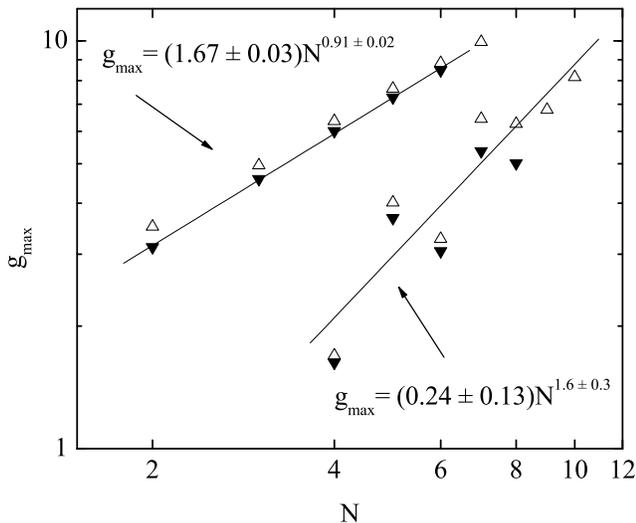}
    \caption{Scaling of $g_{\rm max}$ with number of particles for the
    Pfaffian (lower set) and Laughlin (upper set) states.  The $g_{\rm max}$ values were obtained
    using the method described in Fig.~\ref{gmax}.  Open and filled
    triangles depict the results obtained by using two and three Landau levels, respectively.
    The straight lines are the best fits through the 2LL data for the Pfaffian state and
    3LL data for the Laughlin state.}
    \label{fits_strong} 
\end{figure}

We now proceed to determine $g_{\rm max}(N)$ for various $N$ values,
for each of the four ground states defined in the previous paragraph.
>From Fig.~\ref{fits_weak}, it is apparent that, for the $L=N$ state at least, 3 Landau levels (3LL's) or more
are required to accurately describe the $g_{\rm max}(N)$ for $N \ge 10$ since
the 2LL curve slightly deviates from the power law for these large $N$'s.  While it
would have been advantageous to examine the effect of even higher Landau
levels, computational constraints have prevented us from doing so.
Nevertheless, we are confident that 3LL's are sufficient for our purposes
since we do not observe any significant deviation from a power law scaling for the small
number of particles that we have considered.  Both the $L=N$ and $L=2(N-1)$ states give a scaling relation that is consistent with
Eq.~(\ref{g_scaling_trad}) since we find that $g_{\rm max}\propto N^{-1}$.
Most important, the prefactor for the $L=2(N-1)$ transition is more than an order of magnitude
larger than that for the single vortex state. The numerics indicate that this prefactor continues
to increase for all the weakly correlated ground states (those with angular
momentum $L\propto N$), though we don't have enough data to give quantitative
predictions.

The results for $g_{\rm max}(N)$ are markedly different for the
Pfaffian and Laughlin states.  A cursory glance at
Fig.~\ref{fits_strong} shows that in fact $g_{\rm max}$
\emph{increases} with particle number, which is a drastic
departure from the scaling relation~(\ref{g_scaling_trad}). Again,
the prefactor increases as the angular momentum of these strongly
correlated ground states (where $L\propto N^2$) increases.  The ground state
with $L=N=4$ is believed to be close to the
Pfaffian~\cite{Popp:2004} and is plotted with the other Pfaffian
data, though numerically it appears to be mixed with the
single-vortex state. Our calculations are restricted to $N\leq 10$
for the Pfaffian and $N\leq 7$ for the Laughlin state, and thus it
is difficult to make quantitative predictions based on finite-size
scaling with such few atoms. Despite these limitations, however,
the numerics clearly indicate that the standard criteria for the
validity of the LLL approximation does not apply to highly
correlated ground states.

\begin{figure}[t]
    \includegraphics[width=0.48\textwidth]{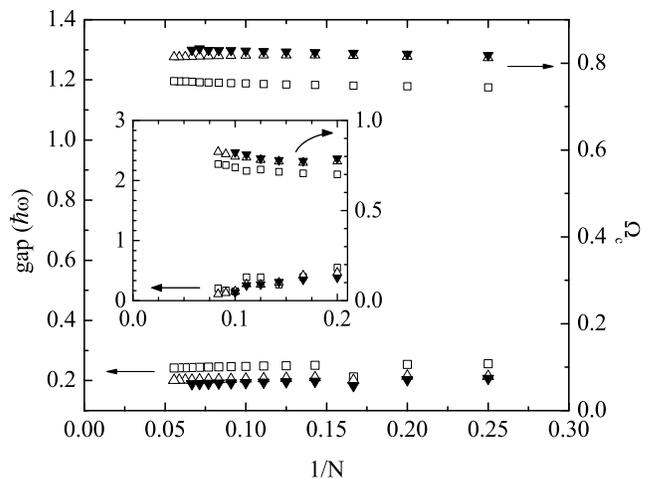} \caption{The critical
    frequency $\Omega_c$ and gap $\Delta E$ are shown as a function of
    particle number $N$ for the $L=N$ state and for the $L=2(N-1)$ state in
    the inset (which has the same axis labels as the main figure,
    suppressed for clarity). For each value of $N$ the coupling constant is
    $g_{\rm max}$ obtained from the best fit to the data shown
    in Fig.~\ref{fits_weak}.}
    \label{scaling_weak}
\end{figure}

Now that we have established the limits on the interaction strength required
to ensure the validity of the LLL approximation (for each value of $N$
considered), we turn our attention to the values of $\Omega_c$ and the
energy gap $\Delta E$ between the ground and first excited state as the number
of particles is increased. The central motivation is to use finite-size scaling
to make predictions for the critical rotation frequencies and gap sizes for
values of $N$ approaching those relevant to future experiments. 

Using the $g_{\rm max}(N)\propto N^{-1.046}$ and $N^{-1.01}$
obtained from the best fits to the data shown in
Fig.~\ref{fits_weak}, the scaling of $\Omega_c$ and $\Delta E$
with particle number is shown in Fig.~\ref{scaling_weak} for the
$L=N$ and $L=2(N-1)$ states. The results for the $L=N$ state are
mostly consistent with expectations in the sense that in the
thermodynamic limit ($N\to\infty$) the critical frequency does
approach a constant value, albeit larger than the mean-field
result ($\approx 0.85$ versus
$\Omega_c=1/\sqrt{2}$~\cite{Sinha:2001,Kasamatsu:2003}), while the
gap goes to a constant $\Delta E\approx 0.2$ in units of
$\hbar\omega$. The asymptotic behavior of the $L=2(N-1)$ state is
more interesting, with the large-$N$ value of $\Omega_c$ appearing
to approach unity while the gap appears to close.

\begin{figure}[t]
    \includegraphics[width=0.48\textwidth]{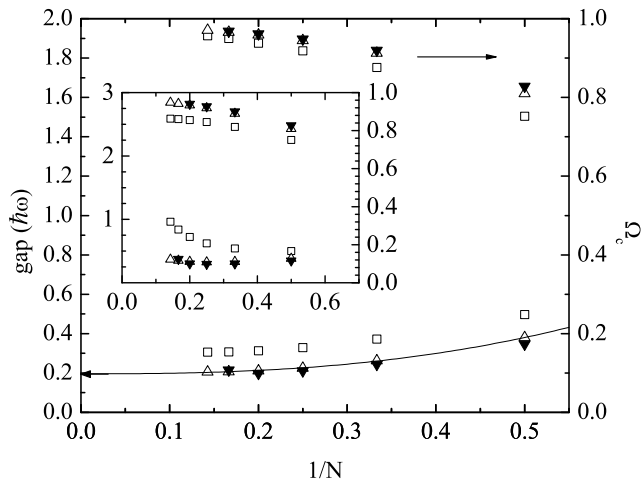}
    \caption{The critical frequency $\Omega_c$ and gap $\Delta E$ are shown as
    a function of particle number $N$ for the Laughlin state. The coupling
    constant $g=3.12$ is the same for all $N$, the value corresponding to the
    largest fixed $g_{\rm max}$ value consistent with the LLL approximation for this state.
    The inset shows the variation of $\Omega_c$ and $\Delta E$ where
    $g_{\rm max}\propto N^{0.91}$ obtained from the best fits to the data shown
    in Fig.~\ref{fits_strong}.}
    \label{scaling_laughlin} 
\end{figure}

\begin{figure}[t]
    \includegraphics[width=0.48\textwidth]{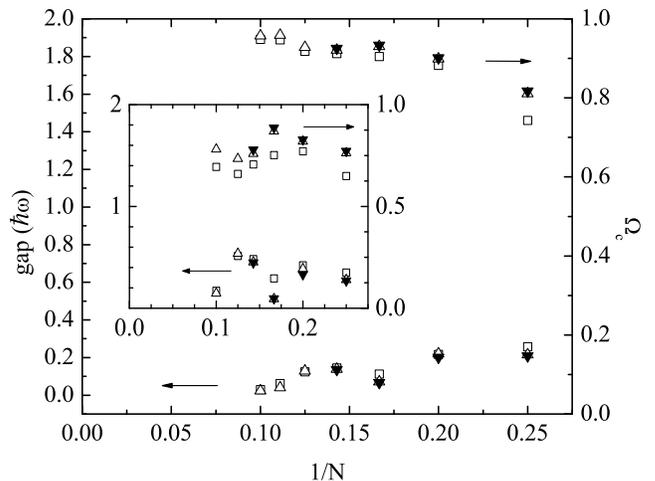}
    \caption{The critical frequency $\Omega_c$ and gap $\Delta E$ are shown as
    a function of particle number $N$ for the Pfaffian. The coupling constant
    $g=1.62$ is the same for all $N$, the value corresponding to the largest constant
    $g_{\rm max}$ consistent with the LLL approximation for this state. The
    inset shows the variation of $\Omega_c$ and $\Delta E$ where
    $g_{\rm max}\propto N^{1.6}$ obtained from the best fits to the data shown
    in Fig.~\ref{fits_strong}.}
    \label{scaling_pfaffian}
\end{figure}

The results for $\Omega_c$ and $\Delta E$ associated with the Pfaffian and
Laughlin states are shown in Figs.~\ref{scaling_laughlin} and \ref{scaling_pfaffian} respectively. The results for both
$g={\rm const.}$ and the best fit values of $g = g_{\rm max}(N)$ are shown.
The constant $g$ is chosen to be the maximum value satisfying the LLL criterion
for all the $N$ in the series (which corresponds to that for the smallest $N$).
The varying $g$ values are the best fits to the data shown in
Fig.~\ref{fits_strong}.

The data shown in Fig.~\ref{scaling_laughlin} clearly indicate
that when $g$ is kept fixed the Laughlin gap approaches a constant
in the large-$N$ limit; this result is consistent with previous
numerical studies of the neutral-atom Laughlin state on the
surface of a sphere~\cite{Regnault:2003}. For $g=g_{\rm
max}(N=2)=3.12$, the asymptotic gap is found to be $\Delta E(N\gg
1)\approx 0.2$ in units of $\hbar\omega$. Unfortunately, when
keeping $g$ fixed the critical rotation frequency approaches the
experimentally challenging limit $\Omega_c\to 1$. On the other
hand, if $g$ increases with $N$ according to its maximum value
$g_{\rm max}(N)$, it appears that the critical frequency levels
off slightly before $\Omega_c=1$ for large $N$ while $\Delta E$ is
larger than in the LLL case, as shown in the inset of
Fig.~\ref{scaling_laughlin}.  Although an exact value is
impossible to obtain, it seems like the gap is approximately
0.4$\hbar\omega$ for large $N$. This implies that with a judicious
choice for $g\gg 1$, the Laughlin state could be stabilized at lower
rotation frequencies without violating the LLL condition.

The data shown in Fig.~\ref{scaling_pfaffian} for the Pfaffian are
less clear than those for the Laughlin state. For fixed
$g=g_{\rm max}(N=4)=1.62$, the critical frequency approaches the high
rotation limit $\Omega_c=1$ while the gap in fact appears to close
for large $N$.  On the other hand, increasing $g$ proportional to
$N^{1.6}$ according to the trend observed in
Fig.~\ref{fits_strong} yields what appears to be a constant value
for the gap and a critical frequency slightly lower than the
centrifugal limit (though there is too much scatter in the data to
make definite statements). Thus, like the Laughlin state, the
Pfaffian can in principle be stabilized at experimentally
accessible rotation frequencies without violating the LLL
condition.

The striking contrast between the weakly and strongly correlated
states in the behavior of $g_{\rm max}(N)$ is the result of a
decreased sensitivity of energy levels to changes in interaction
strength and number of Landau levels, as the total angular
momentum increases. In the LLL approximation, the energy spectrum
is invariant (up to an overall scale factor) under changes in the
interaction strength. The Laughlin state by definition remains a
zero-energy eigenstate of the many-body Hamiltonian. Thus, the
shift with $g$ of all other levels must be directly proportional
to their energies relative to that of the Laughlin state. In
short, the sensitivity to interactions as a function of angular
momentum must be proportional to the $\Omega=1$ `yrast'
line~\cite{Mottelson:1999}. Numerically, we find that the relative
ground-state energies for a given value of $L>N$ decrease
exponentially with angular momentum, $E_{\rm min}\propto
\exp(-\gamma L)$, where $\gamma=0.24$, $0.16$, and $0.08$ for
$N=4$, $5$, and $6$, respectively.  For $2 \le L \le N$ the
relationship is linear
\cite{Jackson:2000,Smith:2000,Papenbrock:2001}.  Small angular
momentum states have thus an inherently increased sensitivity to
changes in the interaction strength.


Low angular momentum states also have an increased sensitivity to
the presence of higher Landau levels.  We quantify this
sensitivity by the resulting shift $\delta E$ in ground-state
energy.  Like $E_{\rm min}$, $\delta E$ decreases with $L$,
although the functional form is different: we find that $\delta
E(L) \propto L^x$ where $x=-1.0$, $-1.2$, $-1.3$ for 4, 5, and 6
particles, respectively. Most relevant though is the scaling of
$\delta E$ with $N$ for the weakly and strongly correlated states.
Recall that for each $N$, $g_\text{max}$ was obtained by setting a
10\% limit on variations of $\Omega_c$ resulting from the
encroaching presence of higher Landau levels. Since $\Omega_c$
characterizes the transition between two states, any variation in
$\Omega_c$ will directly result from a mismatch in $\delta E$
between these two states.  We observe the $L=N$ state to always
transition from $L=0$, and the Laughlin state from $L=N(N-2)$.
After numerically evaluating $\delta E(N)$ with $g=1$ and
$\Omega=1$, we find that $\left[\delta E_{L=0}-\delta
E_{L=N}\right](N) \propto N^{2.8}$ while
$\left[E_{L=N(N-2)}-\delta E_{\rm Laughlin}\right](N) \propto
N^{-0.054}$.  Let the difference of $\delta E$ between two states
be called $\Gamma$.  Should $\Gamma=0$, then both state energies
are equally shifted, and as a result there is no change in the
critical transition rotation frequency $\Omega_c$.  In other
words, $\Gamma$ characterizes the offset in $\Omega_c$ compared to
its LLL value.  As $N$ is increased, any increase in $\Gamma$ can
be countered by a concomitant decrease in the interaction
strength, and vice versa, such that a constant higher Landau level
influence is maintained.  For the $L=0$ to $L=N$ transition $g$
must be decreased with $N$ to keep $\Gamma(N)$ unchanged while for
the Laughlin transition the reverse is true. The contrasting
behavior of $g_{\rm max}(N)$ for the weakly and strongly
correlated states is thus recovered.


Though it is somewhat silly to extrapolate these small-$N$ results to
experimentally relevant values of $N\sim 10^3$, it is nevertheless useful to
approximately determine the relevant experimental parameters required to
stabilize the Laughlin and Pfaffian states. First, it is important that the
excitation gap be larger than the temperature; for a typical temperature
$T\sim 10$~nK one requires that $\Delta\tilde{E}/h\gtrsim 200$~Hz. Thus, stabilizing
the Pfaffian state with an asymptotic gap of $0.1\hbar\omega$ would therefore
require a large radial trap frequency on the order of kHz unless the
temperature were strongly reduced.

The Laughlin state with $\Delta E\approx 0.4$ is more
accessible. This gap size corresponds to $N=7$ and $g\approx 10$
in the data presented in Fig.~\ref{fits_strong}. Assuming that the
$N=7$ result is already close to the large-$N$ asymptotic gap,
which seems to be the case according to the inset of
Fig.~\ref{scaling_laughlin}, then $g=10$ corresponds to a
scattering length of approximately 8~$\mu$m  $\approx 1500a_{\rm Rb}$, assuming
$\omega/2\pi = 200$Hz and $\omega_z = 1$kHz, and where $a_{\rm
Rb}$ is the s-wave scattering length for
$^{87}$Rb~\cite{vanKempen:2002}. Such a large scattering length
can be obtained through the use of Feshbach
resonances~\cite{Tiesinga:1993}, but then the pseudopotential
employed in the present calculations to describe the low-energy
atomic collisions would have to be modified to explicitly take
into account the presence of atomic pairs~\cite{Cooper:2004}.

\section{Attractive Interactions}\label{negative-g}

We also have investigated the effect of having attractive interactions
between bosons.  In the thermodynamic limit, it is known that gases of
untrapped bosons are unstable against collapse~\cite{Nozieres:1982,Stoof:1993}
when $g<0$.  However, with the addition of a trapping potential, such gases
can stably exist for a limited number of particles~\cite{Bradley:1997} because
of the finite zero-point energy of the trap. It has been suggested
recently~\cite{Fischer:2004,Lakhoua:2005} that for the special case of a
rotating, attractive gas of 2D harmonically trapped bosons, there might not be
an upper limit to the number of particles in which a stable gas could exist.
The claim is that the statistical pressure of the anyons comprising the
quantum Hall states can stabilize the cloud against collapse. The argument,
however, has recently been disputed~\cite{Ghosh:2004}.

\begin{figure}[t]
    \includegraphics[width=0.48\textwidth]{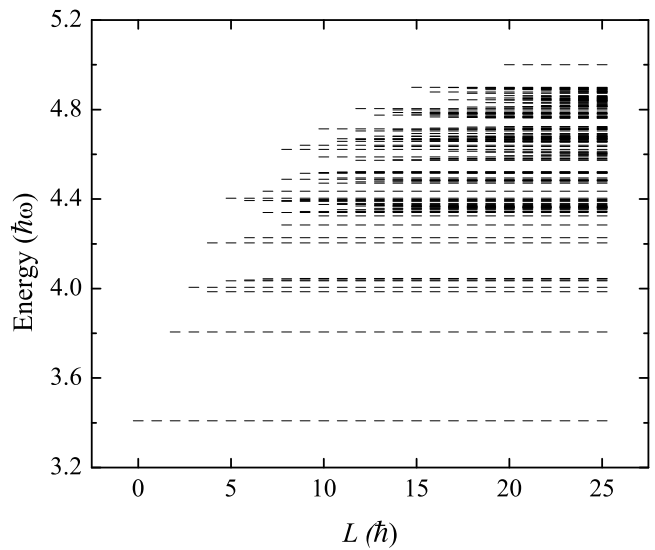} \caption{Spectrum
    of a boson gas of 5 particles with attractive interactions, using the LLL approximation.
    For this particular case, $g=-1.0$ and $\Omega=1$.}
    \label{spectrum_neg}
\end{figure}

The main result of Ref.~\cite{Fischer:2004} is that the bare interaction is
replaced by an effective constant:
\begin{equation*}
g_\text{eff}=g\mp\frac{4\pi}{m\kappa}.
\end{equation*}
Here, $\kappa$ is the Chern-Simons coefficient that relates the
(statistical) magnetic field $B$ to the particle density $\rho$
through $-\kappa B = 4\pi \rho$. In the rotating system,
$B=-2m\Omega/\hbar$ and the effective interaction strength is then
\begin{equation}
g_\text{eff}=g\mp\frac{4\pi\hbar\Omega}{\rho}
\label{g_eff}, 
\end{equation}
and thus for a specific rotation frequency $\Omega$ there exists a
critical negative interaction strength
$g=g_c=-4\pi\hbar\Omega/\rho$ above which the gas can be
stabilized by the so-called anyonic pressure.   To consider the
validity of this claim, we examine the two distinct cases where
$\Omega=\omega$ and $\Omega < \omega$.

\begin{figure}[t]
    \includegraphics[width=0.48\textwidth]{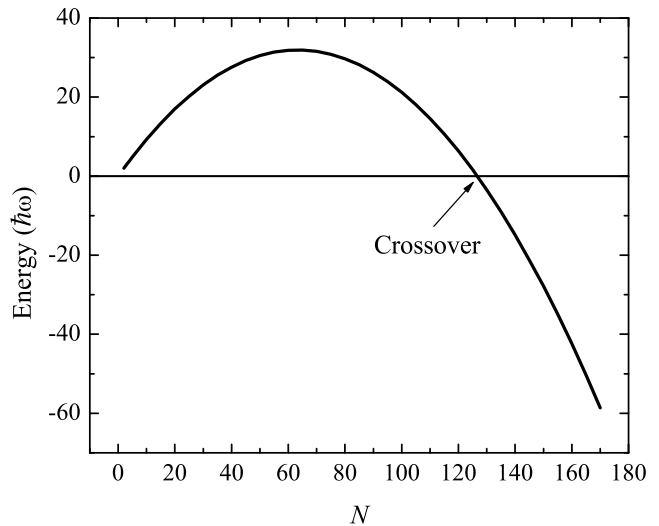} \caption{Ground
    state energy for $g=-0.1$ showing the crossover to instability
    for $N=126$ particles. \label{crossover}}
\end{figure}

Let us begin by considering the high rotation frequency limit $\Omega=\omega$.
In this case, the energy spectrum depends solely on the interaction
Hamiltonian~(\ref{Hint}).  A sample spectrum of the Bose gas when $g<0$ is
shown in Fig.~\ref{spectrum_neg} for $N=5$ and $g=-1.0$. The spectrum is identical to
that for positive interactions, except that it is inverted about the Laughlin
state. The ground state energy is now degenerate with that of the $L=0$ state,
which can be obtained by directly evaluating Eq.~(\ref{Hint-2nd}) with all
particles having $n$, $m=0$:
\begin{equation}
E_{L=0} = N + g\frac{N(N-1)}{4\pi}
\label{EL=0}.
\end{equation}
It is important to emphasize that when $\Omega=\omega$, no matter how much
$L$ is imparted to the system, the ground state energy will always be the same as
for the $L=0$ state.  Thus, we see that from Eq.~\ref{EL=0} the gas is unstable ($E<0$)
when $N>1-(4\pi/g)$. Accordingly, in the thermodynamic
limit, the attractive gas will never be stable
even for the weakest interaction parameter: for large enough $N$ the system always crosses
over into an unstable regime, as seen in Fig.~\ref{crossover}. This result is intuitively
obvious: with attractive interactions, the ground state has $L=0$, which
obviously contains no anyons because it is non-rotating. Thus, the
stabilization observed when $N$ is smaller than its critical value
is due solely to the zero point energy.

If we now turn our attention to the situation where $\Omega <
\omega$, states with $L\neq 0$ have energies that are no longer
degenerate with the $L=0$ state due to the non-zero term
$(1-\Omega/\omega)L$ in the $H_0$ eigenvalue equation.  In this
case, as $L$ is increased, so does its associated ground state
energy. It is therefore plausible that states with large $L$ might
have positive energy. According to this scheme, it would be
preferable to have low $\Omega$ since this would ``lift" the high
$L$ states further away from the bottom of the spectrum---larger
attractive interactions would be stabilized by lower rotation
frequencies. However, this is contradictory to Eq.~(\ref{g_eff})
which states that for fixed $\rho$ (which is the case when $L$ is
kept constant) \emph{increasing} $\Omega$ increases the maximum
attractive interaction strength $g_c$ that can be stabilized by
the anyonic pressure.   In other words, suppose that we decrease
$\Omega$ while  maintaining a constant amount of $L$.  This will
ensure that $\rho$ remains constant and so, according to Eq.~(4),
$|g_c|$ should increase with $\Omega$. However, at fixed  $L$, the
$H_0$ energy is larger for smaller $\Omega$ meaning that the
maximum attractive interaction strength for which the gas is
stable should also be larger. This contradicts the predictions of
Eq.~(4), and thus we have one more reason to refute the claims
brought forth in Ref.~\cite{Fischer:2004}. While it may be
possible to stabilize an attractive boson gas consisting of a
finite number particles in a high $L$ state, the system will
always become instable if $N$ is allowed to increase without
bounds.  Other objections have been raised in the
comment~\cite{Ghosh:2004}.

\section{Conclusions}\label{conclusion}

In summary, we have determined the maximum interaction strength
that can be used before the LLL approximation becomes invalid. For
`mean-field' states with $L\propto N$, we obtain a $g_{\rm max}$
which decreases with $N$. With this scaling, in the limit of large
$N$ both the excitation gap and critical rotation frequency for
the $L=N$ `single-vortex' state approach limiting values of around
$0.2\hbar\omega$ and $0.85\hbar\omega$, respectively. In the same
limit for the $L=2(N-1)$ state, however, the gap closes and the
critical frequency approaches the radial trap frequency. For the
`strongly correlated' states where $L\propto N^2$, we find the
remarkable result that the maximum interaction strength consistent
with the LLL approximation actually \emph{increases}: $g_{\rm
max}\propto N^{x}$, where $x$ is a positive number ($=1.6$ for the
Pfaffian and $0.91$ for the Laughlin states). This in turn implies
that the critical rotation frequencies required to stabilize these
states could be noticeably smaller than the radial trap frequency.
The gaps for the Laughlin and Pfaffian approach values of
0.4$\hbar\omega$ and 0.1$\hbar\omega$ respectively.  On other
hand, if the interaction strength is chosen to be a constant for
all $N$, the numerics show that in the large-$N$ limit the
critical frequency for the `strongly correlated' states approaches
the trap frequency, and that for the Laughlin state the energy gap
approaches a non-zero value of 0.2, while for the Pfaffian it
appears to close.

We also have investigated the regime of negative interaction strengths, and
conclude that in the thermodynamic limit, the trapped Bose gas is unstable
against collapse for any rotation frequency, contrary to previous claims.

\acknowledgments

This work was supported by the Natural Sciences and Engineering
Research Council of Canada, the Canada Foundation for Innovation
and Alberta's Informatics Circle of Research Excellence.

\bibliographystyle{aip}
\bibliography{interactions}

\begin{thebibliography}{10}
\bibitem{Petrov:2001}
This assumes the parameters are chosen so that the scattering states are far
  from resonance with the virtual bound state. See D.~S.~Petrov and
  G.~V.~Shlyapnikov, Phys. Rev. A~{\bf 64}, 012706 (2001).
\end{thebibliography}

\end{document}